\documentclass[12pt,onecolumn,draftcls]{IEEEtran}

\usepackage{graphicx}
\usepackage{amssymb}
\usepackage{verbatim}
\usepackage{graphicx}
\usepackage{amsmath}

\newtheorem{thm}{Theorem}[section]
\newtheorem{cor}[thm]{Corollary}
\newtheorem{lem}[thm]{Lemma}
\newtheorem{defn}[thm]{Definition}

\begin{document}
\title{A Distributed Dynamic Frequency Allocation Algorithm}
\author{Behtash~Babadi and
        Vahid~Tarokh
\thanks{B. Babadi and V. Tarokh are with the School of Engineering and Applied Sciences, Harvard University, Cambridge,
MA, 02138. (e-mails: \{behtash , vahid\}@seas.harvard.edu)}}
\maketitle
\begin{abstract}
We consider a network model where the nodes are grouped into a number of clusters and propose a distributed dynamic frequency allocation algorithm that achieves performance close to that of a centralized optimal algorithm. Each cluster chooses its transmission frequency band based on its knowledge of the interference that it experiences. The convergence of the proposed distributed algorithm to a sub-optimal frequency allocation pattern is proved. For some specific cases of spatial distributions of the clusters in the network, asymptotic bounds on the performance of the algorithm are derived and comparisons to the performance of optimal centralized solutions are made. These analytic results and additional simulation studies verify performance close to that of an optimum centralized frequency allocation algorithm. It is demonstrated that the algorithm achieves about 90$\%$ of the Shannon capacities corresponding to the optimum/near-optimum centralized frequency band assignments. Furthermore, we consider the scenario where each cluster can be in active or inactive mode according to a two-state Markov model. We derive conditions to guarantee finite steady state variance for the
output of the algorithm using stochastic analysis. Further simulation studies confirm the results of stochastic modeling and the performance of the algorithm in the time-varying setup.
\end{abstract}

\section{Introduction}

Dynamic frequency allocation has an important role in the performance of wireless ad-hoc networks, for it results in less transmission power, which is a crucial objective in communication networks. To do this in an optimal way, there needs to be a centralized processor with full knowledge of the spatial distribution profile of the network clusters. However, in many emerging wireless networks (such as ad hoc networks, cognitive radios, etc.), no central frequency allocation authority is naturally available. This makes distributed frequency allocation an important, but mostly unchartered territory in wireless networking.

Centralized frequency allocation has been extensively studied in the literature (Please see \cite{macdonald} and \cite{steele}). There are a number of proposed solutions to similar problems in different contexts (Please see \cite{cendrillon}, \cite{etkin}, \cite{huang}, \cite{peng}, \cite{ramanathan}, and \cite{yu}). These include methods based on graph coloring for cognitive networks, iterative waterfilling for Digital Subscriber Lines (DSL), etc. These approaches either excessively simplify the interference models, or are not fully decentralized, or require too much information exchange between autonomous nodes/clusters, or suffer from all these shortcomings. Additionally, they are all too complex to implement. In \cite{ramanathan}, the approach is based on approximating the optimal resource allocation solutions on a graph. Others \cite{peng} propose that secondary users choose their spectrum according to their information about their local primary and secondary neighbors. They employ a simplified model for mutual interference of the network nodes that turns the problem into the graph multi-coloring problem. They subsequently compute a sub-optimal solution to the graph multi-coloring by using an approximation algorithm to the graph labeling problem.

In the context of Digital Subscriber Lines (DSL), some recent works regarding spectrum balancing have been done (Please see \cite{cendrillon} and \cite{yu}). The objective of spectrum balancing in DSL systems is to maximize the throughput of each user by shaping its Power Spectral Density (PSD) of transmission, satisfying a certain power constraint. In \cite{yu}, a method of iterative waterfilling is proposed in order to solve the problem. In the case of two users, they show the existence and conditions on the uniqueness of a Nash equilibrium point for the iterative algorithm. However, each user must know a weighted sum of the PSD of the other users (interference), in order to do waterfilling. The iterative waterfilling algorithm has high complexity and the resulting Nash equilibrium point is not necessarily the optimal solution. For instance, in a two-user scenario, if both users start with a flat PSD initially, iterative waterfilling does not change their PSD. This is clearly a Nash equilibrium point, but is far away from the optimal answer. In \cite{etkin}, it is shown that this non-optimal Nash equilibrium point might be the only Nash equilibrium, and therefore iterative waterfilling fails for various scenarios.

In \cite{cendrillon}, the users need to balance their power along a number of tones in order to optimize their throughout under power constraints. The optimization problem is relaxed based on introducing a virtual user with fixed thresholds. It turns the problem into a separable optimization problem across the tones for different users. An algorithm has been proposed to solve the relaxed problem iteratively via solving local optimization problems by the users. The knowledge of a weighted sum of the PSD of the other users (interference) is required for each user to solve its local optimization problem. The convergence of the algorithm has been shown in high SNR regime. Simulations show that the achievable region resulted by the solution of the relaxed distributed optimization is close to that of the optimal centralized solution. However, no one-to-one correspondence between the points of the achievable regions of the optimal (centralized) and decentralized algorithms is guaranteed. Therefore, the algorithm does not necessarily converge to optimal values. For the case of asynchronous transmission (in the presence of ICI), the optimization problem is not separable across the tones. They have therefore used heuristic optimization approaches with no convergence guarantees.

In \cite{etkin}, it is shown that the problem of optimal PSD shaping across the users is reducible to that of allocating piece-wise constant powers. This result reduces the complexity of the spectrum sharing problem. Furthermore, a number of achievability and existence results in the context of non-cooperative and cooperative game theory for obtaining efficiency and fairness have been established. Another approach has been presented in \cite{huang}, where each user in the network announces a price to the other users to adapt the power allocation accordingly. Convergence results have been established using supermodular game theory.

In the model we consider, the nodes are divided into different clusters and each cluster is represented with a cluster head. This is motivated by the fact that various networks are naturally clustered (e.g., combat scenario, WLAN Hotspots, WPAN). Each cluster head, having knowledge about the interference it experiences, chooses the frequency band with the least amount of interference from the other clusters. The channel model we consider is the common path loss model, which gives a more refined model than that of the existing literature. It is shown that this distributed strategy converges to a sub-optimal spectrum assignment, without any cross-cluster information exchange. In other words, the algorithm converges to a local minimum of the aggregate interference of the network. Simulation results (Section \ref{simulation_results}) show that the minimization of the aggregate interference of the network results in a sub-optimal solution for the problem of maximizing the aggregate Shannon capacity of the network links. It must be noted that regardless of the model, channel reciprocity is sufficient for the convergence of the algorithm.

The proposed algorithm provides a simple, fully distributed, dynamic frequency allocation strategy that requires neither any information exchange between autonomous devices, nor even any knowledge of the existence of other autonomous entities. Additionally, the proposed algorithm can be used in conjunction with any realistic wireless radio channel model such as those commonly employed in wireless standards (Hata model, Okumura model, etc.). We will further propose performance bounds on this sub-optimal spectrum assignment for some specific network topologies.

We also present a framework to analyze the scenario that clusters can be in sleep or active mode, and go off and on according to time-varying statistics. In our model, the activity of the clusters is described by a stochastic process. For the simplicity of the analysis, we assume that all the clusters go on and off according to a two-state Markov model independently. We have shown that the temporal dynamics of the algorithm can be accurately modeled by an exponential decay. We have derived a stochastic dynamical equation for the evolution of the aggregate interference of the network. The time variation of the activity of clusters is also included in the model, which results in a stochastic differential equation for the steady state behavior of the algorithm. We have further derived a trade-off inequality in terms of the update rate, switching rate between sleep and active mode and the geometrical properties of the distribution of users, guaranteeing a certain steady state variance on the performance of the algorithm. Simulation results verify the accuracy of stochastic modeling and the performance of the algorithm in the presence of time-varying statistics.

The outline of this paper follows next. In Section \ref{system_model}, we discuss the system model and the assumptions employed for our analysis. We disclose the Main Algorithm in Section \ref{the_algorithm}. In Section \ref{main_results}, we present the main results of this paper regarding the convergence and performance of the main algorithm. Simulation results for some specific network topologies are provided in Section \ref{simulation_results}. Finally, we discuss the contributions of the proposed method and future areas of research in Section \ref{conclusion}. The proofs of main results can be found in Appendices \ref{convergence} and \ref{proofs}.

\section{System Model}\label{system_model}
Suppose that we have a set of network nodes distributed in space such that they can be partitioned into a union of clusters. These networks often happen in nature, for instance, in a combat scenario a group of soldiers can be divided into a number of clusters according to their missions. Communication within clusters is then very desired. There are a number of efficient methods for partitioning the network elements, but this topic is not the focus of this paper. We assume that the clusters are already formed in a specified manner.

In light of the above, our network model is given by collection of nodes in $N$ clusters, $c_i$, $i=1,\cdots,N$, where each cluster has a cluster head responsible for managing some of the network functions. Let $d_{ij}$ denote the distance between the cluster heads of $c_i$ and $c_j$ (Fig. 1). We use the following assumptions for the model:

$\bullet$ The $i$th cluster, $c_i$, contains $n_i$ users.

$\bullet$ At each time slot for any cluster \emph{at most} one user is transmitting and one user is receiving. For simplicity of mathematical analysis, we have assumed that at each time slot for any cluster \emph{exactly} one user is transmitting and one user is receiving (See Sections \ref{extension} and \ref{simulation_results}). This assumption can be relaxed to any scenario satisfying channel reciprocity between clusters. For instance, in an alternative scenario each user transmits and receives its data through the cluster head. This model also satisfies the channel reciprocity conditions. Therefore, the results can be generalized to various other models.

$\bullet$ Each user transmits with power $KP_0$, where $K$ is a constant such that the power at $1$ meter is $P_0$. The assumption of equal transmission powers can be further relaxed and is adopted for mathematical convenience.

$\bullet$ The distances between clusters are much larger than the size of clusters and bounded below by a distance $\delta$.

$\bullet$ The transmission model is path loss with exponent $\eta$. No shadowing and fading is assumed in the current analysis. However, the analysis can be generalized to more realistic models of transmission.

$\bullet$ The accessible spectrum is divided into $r$ different bands, denoted by $b_1,\cdots,b_r$.

$\bullet$ At time $t$, the $i$th cluster is in state $s_i(t) \in \{1,2,\cdots,r\}$, corresponding to the index of the transmission band it is using.

$\bullet$ The probability of two clusters updating their frequency bands at the same instance of time is negligible. This assumption can be relaxed \cite{collision}.

$\bullet$ The rate of change of the spatial distributions of the clusters in the network is much less than the processing/transmission rate. Therefore, the topology of the network is assumed to be fixed in the analysis of the frequency allocation algorithm.

$\bullet$ We choose the aggregate interference of the network as our performance metric. This assumptions makes the analysis mathematically tractable and is a reasonable metric for performance comparison. Simulation results in Section \ref{simulation_results} show that the minimization of the aggregate interference of the network results in a near-optimal solution for the problem of maximizing the aggregate Shannon capacity of the network links.

\section{The Algorithm}\label{the_algorithm}

Using the above assumptions, we approximate the $i$th cluster, $c_i$, by a single node with transmission power $P_0$ within a distance $d_{ij}$ from the other clusters. The interference experienced by $c_i$ caused by all the other clusters is therefore,
\begin{equation}\label{interference}
I_{c_i}(N,\{d_{ij}\},t)=\sum_{j \ne i}\frac{P_0}{d_{ij}^n} \delta(s_i(t),s_j(t))
\end{equation}
where $\delta(x,y)$ is the Kronecker delta function, defined as
\begin{equation} \label{delta}
\delta(x,y)= \left\{ {\begin{array}{*{20}c}
   {1} &  &{x=y}  \\
   {0} &  &{x\neq y}  \\
\end{array}} \right.
\end{equation}

At time $t=t_n$, one of the clusters, say $c_i$, updates its transmission frequency band. The nature of the update procedure is asynchronous for all the clusters. This is intuitively appealing, because of the nature of ad-hoc networks, where there is usually no common clock among the nodes. We assume that the updates are taking place at times $t_1,t_2,\cdots$. We can therefore change the continuous-time interference model in Eq. (\ref{interference}) to a discrete-time version as
\begin{equation}\label{interference_d}
I_{c_i}(N,\{d_{ij}\},l)=\sum_{j \ne i}\frac{P_0}{d_{ij}^{\eta}} \delta(s_i(l),s_j(l))
\end{equation}
where $l$ corresponds to the time $t=t_l$, when an update is taking place. Let $C_k(N,\{d_{ij}\},l)$ denote the set of clusters transmitting in band $b_k$ prior to time $l$. Also, let $I_{c_i,k}(N,\{d_{ij}\},l)$ denote the interference experienced by $c_i$ caused by all the clusters in $C_k(N,\{d_{ij}\},l)$ if $c_i$ was transmitting in band $b_k$, for $k=0,\cdots,r-1$. $I_{c_i,k}(N,\{d_{ij}\},l)$ can be written as
\begin{equation}
I_{c_i,k}(N,\{d_{ij}\},l)=\sum_{c_j \in C_k(N,\{d_{ij}\},l), j \neq i} \frac{P_0}{d_{ij}^{\eta}}
\end{equation}

We denote the aggregate interference of the network at time $l$ by $I(N,\{d_{ij}\},l)$ as
\begin{equation}
I(N,\{d_{ij}\},l) = \sum_{i} I_{c_i}(N,\{d_{ij}\},l) = \sum_{i} \sum_{j \ne i}\frac{P_0}{d_{ij}^{\eta}} \delta(s_i(l),s_j(l))
\end{equation}

It must be noted that for notational convenience, we drop the time dependence of the functions $I(N,\{d_{ij}\},l)$, $I_{c_i}(N,\{d_{ij}\},l)$ and $I_{c_i,k}(N,\{d_{ij}\},l)$ following the convergence of the algorithm or whenever the spatial configuration is fixed over time, and denote them by $I(N,\{d_{ij}\})$, $I_{c_i}(N,\{d_{ij}\})$ and $I_{c_i,k}(N,\{d_{ij}\})$, respectively. We also denote the aggregate interference of the worst-case scenario, optimal scenario and that of the output of the algorithm by $I_w(N,\{d_{ij}\})$, $I_o(N,\{d_{ij}\})$ and $I_a(N,\{d_{ij}\})$, respectively. We can now define the main algorithm:

\textbf{Main Algorithm}: \emph{Clusters scan all the frequency bands $b_1,\cdots,b_r$ in an asynchronous manner over time. Each cluster chooses the frequency band in which it experiences the least aggregate interference from other clusters. In other words, at time $l$, a cluster, say $c_i$, updates its state according to the following rule}
\begin{equation}\label{decision}
s_i(l+1)=\arg \min_k I_{c_i,k}(N,\{d_{ij}\},l)
\end{equation}
\emph{where $s_i(l+1)$ is the new state of $c_i$ updated at time $l$.}

For this purpose, the cluster head scans all the frequency bands and estimates/measures the interference it experiences in each frequency band. The cluster head chooses the new transmission frequency band according to Eq. (\ref{decision}), the decision criterion in the Main Algorithm.

\section{Main Results}\label{main_results}

\subsection{Convergence}
\begin{thm}\label{main}
Given any reciprocal channel model, the Main Algorithm converges to a local minimum in polynomial time in $N$.
\end{thm}

\begin{proof}
The proof is given in Appendix \ref{convergence}.
\end{proof}
\subsection{Performance Bounds}\label{perf_bounds}
\begin{thm}[Upper Bound]\label{upper}
Let $I_a(N,\{d_{ij}\})$ denote the aggregate interference of all the clusters corresponding to the state of the algorithm following convergence (see Theorem \ref{main}), and $I_w(N,\{d_{ij}\})$ to be the aggregate interference for the worst case interference scenario (all clusters transmitting in one frequency band), then
\begin{center}
$I_a(N,\{d_{ij}\}) \le \frac{1}{r} I_w(N,\{d_{ij}\})$
\end{center}
\end{thm}

\begin{proof}
The proof is given in Appendix \ref{perf}.
\end{proof}

In order to obtain further performance bounds for the Main Algorithm, one must analyze its behavior following convergence, over a configuration induced by the spatial configuration of the network nodes. However, such an analysis for the general case is non-trivial. We therefore focus our attention on a specific class of spatial configurations, where all the clusters are co-linear (lie on a line). This is referred to as a \emph{Linear Array}. For a given $N$, we assume that the clusters are located in $[0,(N-1)d]$, where $d$ is a constant.

\begin{defn}
A \emph{Uniform Linear Array} of $N$ clusters is a Linear Array in which $d_{i,i+k}=kd$, for $1 \le i,i+k \le N$.
\end{defn}

\begin{thm}[Optimal Strategy]\label{optimal_strategy}
Let $I_o(N,\{d_{ij}\})$ be the aggregate interference of the optimal strategy for a given linear array of clusters located in $[0,(N-1)d]$. Then,
\begin{equation}\label{b1}
\lim_{N \rightarrow \infty} \frac{1}{N} I_o(N,\{d_{ij}\}) \ge \frac{1}{r^{\eta}} 2 \zeta(\eta) \frac{P_0}{d^{\eta}}
\end{equation}
where $\zeta(\eta)$ is the Riemann zeta function.
\end{thm}

\begin{proof}
The proof is given in Appendix \ref{optimal}.
\end{proof}

\begin{thm}\label{regular}
Uniform Linear Array achieves the bound in the statement of Theorem \ref{optimal_strategy}, as $N \rightarrow \infty$.
\end{thm}

\begin{proof}
The proof is given in Appendix \ref{regular_a}.
\end{proof}

\begin{thm}\label{r=2}
If $r=2$ and $\eta \ge 2$ in the statement of Theorem \ref{regular}, then the optimal strategy is the alternating assignment for any $N$.
\end{thm}

\begin{proof}
The proof is given in Appendix \ref{optimal_2}.
\end{proof}

We can use Theorems \ref{optimal_strategy} and \ref{regular} to upper bound the performance of the algorithm as given by the following Corollaries:

\begin{cor}\label{c1}
Let $I_a(N,\{d_{ij}\})$ denote the aggregate interference corresponding to the output of the Main Algorithm and $I_o(N,\{d_{ij}\})$ be the aggregate interference of the optimal strategy. Then, for any spatial configuration of clusters in $[0,(N-1)d]$ we have
\begin{equation}
\frac{I_a(N,\{d_{ij}\})}{I_o(N,\{d_{ij}\})} \le \frac{r^{\eta-1}}{\big(\frac{d_{min}}{\min\{d_{max},d\}}\big)^{\eta}}
\end{equation}
as $N \rightarrow \infty$.
\end{cor}

\begin{proof}
The proof is given in Appendix \ref{bound_1}.
\end{proof}

\begin{cor}\label{c2}
If the array in the statement of Corollary \ref{c1} is a Uniform Linear Array, then

\[
\frac{I_a(N,d)}{I_o(N,d)} \le r^{\eta-1}
\]
where $I_a(N,d) \triangleq I_a(N,\{d_{ij}\})$ and $I_o(N,d) \triangleq I_o(N,\{d_{ij}\})$ for a Uniform Linear Array in $[0,(N-1)d]$.
\end{cor}
\begin{proof}
The statement of the Corollary follows by letting $d_{max}=d_{min}=d$ in Corollary \ref{c1}.
\end{proof}

\subsection{Time-varying Setup}\label{time_varying}

For cluster $c_i$, $i=1,2,\cdots,N$, we consider an activity indicator state $a_i(l)$, such that $a_i(l)=1$ and $a_i(l)=0$ correspond to being active and inactive at time $l$, respectively. Let $P_0^{c_i}(l)$ and $P_1^{c_i}(l)$ be the probability of $c_i$ being in activity indicator state $0$ and $1$ at time $l$, respectively. The evolution of the probabilities is given by:
\begin{equation}\label{markov}
\left( {\begin{array}{*{20}c}
   {P_0 ^{c_i } (l + 1)}  \\
   {P_1 ^{c_i } (l + 1)}  \\
\end{array}} \right) = \left( {\begin{array}{*{20}c}
   \alpha  & {1 - \alpha }  \\
   {1 - \alpha } & \alpha   \\
\end{array}} \right)\left( {\begin{array}{*{20}c}
   {P_0 ^{c_i } (l)}  \\
   {P_1 ^{c_i } (l)}  \\
\end{array}} \right)
\end{equation}
which corresponds to a symmetric two-state Markov model. We assume $P_1^{c_i}(0)=1$ for all $c_i$, $i=1,\cdots,N$.

In this case, the convergence of the algorithm is not guaranteed. Furthermore, the upper bound on the performance of the algorithm does not hold in presence of time-varying statistics. In order to further analyze the performance of the algorithm, we use the following additional assumptions:

$\bullet$ The clusters update their frequency band asynchronously according to the same temporal statistics.

$\bullet$ The update process is modeled by a Poisson process of rate $\frac{1}{\Delta T}$, \emph{i.e.}, each cluster updates its frequency band with a rate $\frac{1}{N \Delta T}$.

$\bullet$ The number of users switching on/off in each time slot is much less than the total number of users (near equilibrium scenario).

\subsection{Dynamics of The Algorithm}\label{dynamics}

Let $I(t)$ denote a continuous-time approximation to the aggregate interference of the network at time $t$. It can be shown that

\begin{equation}\label{dif_eq_1}
\frac{d \mathcal{E}[{I}(t)]}{dt} = -\frac{\rho}{\tau} (\mathcal{E} [I(t)] - \mathcal{E}[I_a])
\end{equation}
where $\mathcal{E}$ denotes the ensemble average (over different update patterns), $\tau \triangleq N \Delta T$ and $\rho$ is geometrical constant showing the effective number of interacting neighbors to a cluster including itself (the details are given in Appendix \ref{dynamic1}).

Eq. (\ref{dif_eq_1}) is an approximate differential equation describing the behavior of the algorithm near equilibrium. In other words, the aggregate interference decreases exponentially with rate $\rho/\tau$ near the equilibrium point.

We can model the change in the number of active clusters by two Poisson counters of rate $\lambda$ \cite{feller}. We associate two independent Poisson counters $dN_+$ and $dN_-$ to the users going on and off, respectively. We have
\begin{equation}
E(dN_+)=E(dN_-)=\lambda dt
\end{equation}

Each cluster when activated, approximately experiences the instantaneous normalized aggregate interference of the network, due to spatial ergodicity.
If we define $\mathcal{I}(t) \triangleq \mathcal{E} [I(t)]$ and $\mathcal{I}_a(t) \triangleq \mathcal{E} [I_a(t)]$, under the assumption of $\lambda$ being small compared to $\frac{1}{\tau}$, we have the new dynamics in the It\^{o} form \cite{Ito} as

\begin{equation}\label{ito}
d \mathcal{I}(t) = -\frac{\rho}{\tau} (\mathcal{I}(t) - \mathcal{I}_a(t))dt + \frac{4}{N} \mathcal{I}(t) (dN_+ - dN_-)
\end{equation}

Under this model, in the steady state the variance settles down to
\begin{equation}
\sigma_{ss}^2 \triangleq E\Big[\big(\mathcal{I}(t) - E[\mathcal{I}(t)]\big)^2\Big] = \mathcal{I}^2_a \frac{16 \lambda \tau}{N^2 \rho - 16 \lambda \tau}
\end{equation}
given
\begin{equation}\label{inequality}
\frac{16 \lambda \tau}{N^2 \rho} < 1
\end{equation}
(please see Appendix \ref{dynamic2} for details).

According to our model for cluster activities give by Eq. (\ref{markov}), $\lambda = N^2(1-\alpha)/2\tau$. Thus, we get the following trade-off inequality
\begin{equation}
\frac{8(1-\alpha)}{\rho} < 1
\end{equation}
in order to have a finite variance in the steady state.

\subsection{Discussion of the Results}

Theorem \ref{main}, guarantees the convergence of the Main Algorithm in polynomial time in $N$, regardless of the configuration of the nodes in the network. The first performance result is stated in Theorem \ref{upper}, which gives an upper bound on the performance of the algorithm. This result is independent of the topology of the network and is a direct consequence of the structure of the Main Algorithm.

In order to obtain further performance bounds for the Main Algorithm, as stated earlier, we have chosen the class of Linear Arrays for our analysis. Theorem \ref{optimal_strategy} is the main result on the asymptotic performance of the Main Algorithm for Linear Arrays. The Riemann zeta function in the statement of Theorem is merely a consequence of the path loss model for the transmission and the fact that at each time slot, only one transmitter and one receiver are active in each cluster. However, the achievability of the bound in Theorem \ref{optimal_strategy} is not trivial, since the optimal strategy of frequency allocation is not known for a general spatial configuration of the clusters. Theorem \ref{regular} proves the asymptotic achievability of the bound in the statement of Theorem \ref{optimal_strategy} for Uniform Linear Arrays. Theorem \ref{regular}, states that the alternating frequency band allocation achieves the bound in Theorem \ref{optimal_strategy} as $N \rightarrow \infty$, and is therefore the optimal strategy of frequency allocation. In order to analyze the optimal strategy for finite $N$, we focus our attention on the case of $r=2$, \emph{i.e.}, two frequency bands. Theorem \ref{r=2}, states that the optimal strategy is the alternating frequency allocation for finite $N$, when $r=2$.

Corollaries \ref{c1} and \ref{c2} compare the performance of the Main Algorithm with respectively that of the optimal strategy for Linear Arrays and Uniform Linear Arrays. This is done by combining the upper bound of Theorem \ref{upper} with the lower bound of Theorem \ref{optimal_strategy}. The bound in Corollary \ref{c1} is not as tight as the original bounds, since it is worst-case and applicable to any spatial configuration of clusters. As the simulation results show in next section, the algorithm performs significantly better than these bounds.

The stochastic analysis for the time-varying case results in a simple trade-off inequality for design purposes. In other words, for a given time-varying statistics, we can design the update rate to guarantee a finite steady state variance.

The geometrical parameter $\rho$ can be empirically estimated for different network topologies. However, theoretical estimates are possible. For instance, for a uniform linear array of clusters, each cluster has two effective neighbors. Therefore, an estimate of $\rho = 3$ seems reasonable and is verified by simulation results (Please see Section \ref{simulation_results}). If $\lambda = O(N^{1-\epsilon})$ for some $\epsilon > 0$, as $N \rightarrow \infty$, the inequality (\ref{inequality}) always holds. Therefore, the algorithm converges in both mean and variance in the sub-linear regime.

\subsection{Extension of the Results}\label{extension}

Generalization of the result of Theorem \ref{r=2} to $r>2$ is not straightforward, since the combinatorial possibilities of the assignments grow exponentially with $r$. Furthermore, generalization of all the above results to higher dimensions is a non-trivial problem. The reason is that in dimensions greater than 1, the degrees of freedom for the cluster interactions increase dramatically. We are currently studying this scenario extensively.

As noted earlier, any model with channel reciprocity suffices for the convergence of the Main Algorithm (See the proof in Appendix \ref{convergence}). Moreover, the upper bound on the performance of the algorithm given in Theorem \ref{upper} holds for any model with channel reciprocity. The other performance bounds, in their current format, rely on the specific path loss model. However, it may be possible to generalize the same method of analysis to other reciprocal channel models.

The stochastic analysis can be generalized to other statistical models for the activity of the clusters over time. The generalization of the results to $r>2$ is straightforward.

\section{Simulation Results}\label{simulation_results}

Fig. 2 shows the performance of the algorithm on different arrays of 100 clusters in one and two dimensions. Fig. 2 (a), (b) and (c) show the normalized aggregate Shannon capacity (both the optimum/near-optimum value and that of the output of the algorithm) and the normalized aggregate interference of the network as a function of time for a uniform linear array (with $r=2$), integer $\mathbb{Z}_2$ (rectangular) lattice (with $r=4$) and a hexagonal lattice (with $r=4$), respectively (Please see \cite{sloane}). The normalized aggregate Shannon capacity is defined as the aggregate Shannon capacity divided by the number of clusters. Similarly, the normalized aggregate interference is defined as the aggregate interference divided by the number of clusters. Here $d=1$, $P_0=1$ and $\eta=2$. For the rectangular and hexagonal lattices, the computation of the optimum frequency band assignment is very complicated and finding it by exhaustive search was beyond the capabilities of our simulation platforms. Instead, we have compared the performance of the algorithm to that of the $1:4$ frequency reuse pattern as a near-optimal candidate. As it can be observed from the figure, the minimization of the aggregate interference results in an overall increasing behavior of the capacity. In all cases more than 90$\%$ of the capacity of the optimal (near-optimal) centralized frequency assignment is achieved.

Fig. 3 shows the performance of the algorithm on a uniform linear array along with the lower and upper bounds we have obtained in Section \ref{perf_bounds}. Here $r=2$, $d=1$, $P_0=1$ and $\eta=2$. For the initial condition of the algorithm, we let all the clusters to be in frequency band $b_1$. The updates are repeated until the convergence is achieved. As we observe from the figure, the algorithm performs significantly better than the upper bound we have obtained and is less than $1$ dB away from the alternating assignment and the lower bound.

Fig. 4 (a) and (b) show the performance of the Main Algorithm, the worst case and the $1:4$ frequency reuse pattern for rectangular and hexagonal arrays of clusters, respectively. Here $d=1$, $r=4$, $P_0=1$ and $\eta=2$. Although we have not established any lower performance bounds for two-dimensional arrays, the algorithm performs very closely to the centralized 1:4 frequency reuse solution.

In Fig. 5, the dynamics of the algorithm for a uniform linear array of 100 clusters is shown. Here $\alpha=1$ which corresponds to the case of no time-varying statistics. The empirical curve is averaged over 500 different update patterns. The theoretical curve corresponds to an exponential with rate $-\rho/\tau$ for $\rho=3$ and $\tau=1$. As it can be observed from the figure, the theoretical estimate matches the empirical curve very well.

Fig. 6 shows the theoretical and empirical steady state variance of the aggregate interference vs. switching rate ($1-\alpha$) for a uniform linear array of 100 clusters. The empirical curve is obtained by averaging over 500 different realizations of the update process. As it can be observed from the figure, the theoretical variance (with $\rho = 3$) matches the empirical variance.

\section{Conclusion}\label{conclusion}
We have proposed a distributed algorithm for finding a sub-optimal frequency band allocation to the clusters in a network. We have also derived some performance bounds for the special case of linear arrays of clusters. Simulations prove that the algorithm performs significantly better than the performance bounds we have established. We have also derived a stochastic differential equation describing the behavior of the algorithm in presence of time-variation in the activity of clusters near the equilibrium point. A trade-off inequality to guarantee stability in the performance of the algorithm is established. The stochastic modeling framework opens the possibilities of both open loop and closed loop stochastic control. These problems are currently being studied.

\appendices
\section{Proof Of the Convergence Result}\label{convergence}

First we prove the following Lemma:
\begin{lem}\label{Lyapunov}
$I(N,\{d_{ij}\},l)$ is a non-increasing function of $l$ that is bounded from below.
\end{lem}

\begin{proof}
First of all, we need to show that $I(N,\{d_{ij}\},l)$ has a lower bound. We know that the aggregate interference of all the clusters
is a non-negative quantity. Therefore, $I(N,\{d_{ij}\},l) \ge 0$ for all $l$.

Secondly, we need to show that $I(N,\{d_{ij}\},l)$ is a non-increasing function of $l$. Without loss of generality, we assume that $c_i$ has been transmitting in band $b_j$. $I(N,\{d_{ij}\},l)$ can be written as
\begin{equation}
I(N,\{d_{ij}\},l)=\sum_k \sum_{c_m \in C_k, m \neq i} I_{c_m,k}(N,\{d_{ij}\},l)+ 2I_{c_i,j}(N,\{d_{ij}\},l)
\end{equation}
where we have used the channel reciprocity. After the update, the algorithm implies that $c_i$ chooses the new band $b_k$ according to the decision criterion (\ref{decision}) in the statement of the Main Theorem. Therefore,
\begin{eqnarray}
I(N,\{d_{ij}\},l+1) &=& \sum_k \sum_{c_m \in C_k, m \neq i} I_{c_m,k}(N,\{d_{ij}\},l)+ 2I_{c_i,k}(N,\{d_{ij}\},l)\\
\nonumber &=& I(N,\{d_{ij}\},l) + 2\big(I_{c_i,k}(N,\{d_{ij}\},l) - I_{c_i,j}(N,\{d_{ij}\},l)\big)
\end{eqnarray}
where $k$ satisfies the decision criterion (\ref{decision}). According to Eq. (\ref{decision}), we have $I_{c_i,k}(N,\{d_{ij}\},l) \le I_{c_i,j}(N,\{d_{ij}\},l)$, for all $j \neq k$. Therefore,
\begin{equation}
I(N,\{d_{ij}\},l+1) - I(N,\{d_{ij}\},l) = 2\big(I_{c_i,k}(N,\{d_{ij}\},l) - I_{c_i,j}(N,\{d_{ij}\},l)\big) \le 0
\end{equation}
which gives the statement of the Lemma. It must be noted that channel reciprocity is sufficient for this proof to hold.
\end{proof}
\emph{Proof of Theorem \ref{main}:}\\
According to Lemma \ref{Lyapunov}, $I(N,\{d_{ij}\},l)$ is a lower bounded non-increasing function of $l$. It can take at most $r^N$ distinct values, corresponding to the $r^N$ different frequency band assignments to the clusters. Therefore, $\exists l_0$ such that $\forall l>l_0$ we have
$I(N,\{d_{ij}\},l+1)=I(N,\{d_{ij}\},l)$ and no cluster updates its state up to an isomorphism from $\{0,1,\cdots,r-1\}$ to itself.

After each update, the change in $I(N,\{d_{ij}\},l)$ is at least of order $\frac{1}{(N{d_{max}})^{\eta}}$, where $d_{max} = \max_{i,j} \{d_{ij}\}$. The aggregate interference of all the clusters is of order $N$. Therefore, we need at most $O(N^{\eta+1})$ switches to reach the final configuration of the algorithm. After any round of updates, during which all the clusters have updated their state, at least one cluster changes its frequency band. Therefore, the total number of updates is $O(N^{\eta+2})$, which is bounded by a polynomial in $N$. Therefore, the Main Algorithm converges to a local minimum in polynomial time in $N$.

\section{Performance Bounds Proofs}\label{proofs}

\subsection{Proof of Theorem \ref{upper}}\label{perf}

\begin{lem}\label{lower1}
Let $I_i(N,\{d_{ij}\}) \triangleq {\sum_{j \ne i}\frac{P_0}{d_{ij}^{\eta}}}$ be the interference on $c_i$ when all the other clusters are co-band with it. Then,
\begin{center}
$I_{c_i,k}(N,\{d_{ij}\},l) \le \frac{1}{r} I_i(N,\{d_{ij}\})$
\end{center}
\end{lem}

\begin{proof}
At time $l$, $c_i$ chooses a frequency band, say $k \in \{1,2,\cdots,r\}$ such that $I_{c_i,k}(N,\{d_{ij}\},l) \le I_{c_i,j}(N,\{d_{ij}\},l)$, for all $j \neq k$. Therefore,
\begin{equation}
r I_{c_i,k}(N,\{d_{ij}\},l) \le \sum_j I_{c_i,j}(N,\{d_{ij}\},l) = I_i(N,\{d_{ij}\})
\end{equation}
which gives the statement of the Lemma.
\end{proof}
\emph{Proof of Theorem \ref{upper}}:\\
Let $I_{c_i}(N,\{d_{ij}\})$ denote the value of
$I_{c_i,k}(N,\{d_{ij}\},l)$ following convergence. Using the result of Lemma \ref{lower1}, the aggregate interference can
be written as
\begin{eqnarray}
I_a(N,\{d_{ij}\}) = \sum_i I_{c_i}(N,\{d_{ij}\}) \le \frac{1}{r} \sum_i I_i(N,\{d_{ij}\}) = \frac{1}{r} I_w(N,\{d_{ij}\})
\end{eqnarray}
which gives the statement of the Theorem.
\subsection{Proof of Theorem \ref{optimal_strategy}}\label{optimal}
\begin{lem}\label{zeta}
Let $I_w(N,d)$ denote the aggregate interference of the worst case scenario for a uniform linear array of $N$ clusters in $[0,(N-1)d]$. As $N \rightarrow \infty$, we have
\begin{equation}
\frac{1}{N} I_w(N,d) \rightarrow 2 \zeta(\eta) \frac{P_0}{d^{\eta}}
\end{equation}
where $\zeta(\cdot)$ is the Riemann zeta function.
\end{lem}

\begin{proof}
We have
\begin{eqnarray}
\frac{1}{N} I_w(N,d) &=& \frac{1}{N} \frac{P_0}{d^{\eta}} \sum_{i=1}^N \sum_{j=1, j \ne i}^N \frac{1}{|i-j|^{\eta}}\\
\nonumber &<& \frac{1}{N} \frac{P_0}{d^{\eta}} \sum_{i=1}^N 2 \sum_{j=1}^\infty \frac{1}{j^{\eta}} = 2 \zeta(\eta) \frac{P_0}{d^{\eta}}
\end{eqnarray}

Let $I_i(N,d) \triangleq I_i(N,\{d_{ij}\})$ denote the interference experienced by the cluster $c_i$ located at $(i-1)d$, when all the other clusters are co-band with it, for a uniform linear array of $N$ clusters in $[0,(N-1)d]$. We can write
\begin{equation}
I_i(N,d) = \frac{P_0}{d^{\eta}} \sum_{j \ne i} \frac{1}{|i-j|^{\eta}}
\end{equation}

We need to show that for any small enough $\epsilon > 0$, $\exists N_0$ such that for $N>N_0$, $2 \zeta(\eta) \frac{P_0}{d^{\eta}} - \frac{1}{N} I_w(N,d) < \epsilon \frac{P_0}{d^{\eta}}$. Let's take $\tilde{N}$ such that $\sum_{i=1}^\infty \frac{1}{(\tilde{N}+i)^{\eta}} < \frac{\epsilon}{4}$. For $N > 2(\tilde{N}+1)$ we have
\begin{eqnarray}
\frac{1}{N} I_w(N,d) = \frac{1}{N} \sum_{i=1}^{N} I_i(N,d) \ge \frac{1}{N} \sum_{i=\tilde{N}+1}^{N-\tilde{N}-1} I_i(N,d)
\end{eqnarray}

Clearly for all $\tilde{N} < i < N - \tilde{N}$,
\begin{eqnarray}
2 \zeta(\eta) \frac{P_0}{d^{\eta}} - I_i(N,d) < 2 \frac{P_0}{d^{\eta}} \sum_{i=1}^{\infty} \frac{1}{(\tilde{N}+i)^{\eta}} < \frac{P_0}{d^{\eta}} \frac{\epsilon}{2}
\end{eqnarray}

Therefore,
\begin{eqnarray}
\frac{1}{N} I_w(N,d) > \frac{1}{N} \sum_{i=\tilde{N}+1}^{N-\tilde{N}-1} I_i(N,d) > \frac{1}{N} \frac{P_0}{d^{\eta}}  \textstyle \sum_{i=\tilde{N}+1}^{N-\tilde{N}-1} (2 \zeta(\eta) - \frac{\epsilon}{2}) = \displaystyle \frac{P_0}{d^{\eta}} \textstyle \frac{N-2(\tilde{N}+1)}{N} (2 \zeta(\eta) - \frac{\epsilon}{2})
\end{eqnarray}

If we choose $N_0$ large enough so that $4 \zeta(\eta) \frac{\tilde{N}+1}{N_0} < \epsilon/2$, for all $N > N_0$ we will have
\begin{eqnarray}
\frac{N-2(\tilde{N}+1)}{N} \frac{\epsilon}{2} + 4 \zeta(\eta) \frac{\tilde{N}+1}{N} &<& \epsilon/2 + \epsilon/2 = \epsilon
\end{eqnarray}
which proves the statement of the Lemma.
\end{proof}

\begin{lem}\label{equilibrium}
For a given distribution of clusters in $[0,(N-1)d]$, we have
\begin{eqnarray}
\nonumber \frac{I_w(N,\{d_{ij}\})}{I_w(N,d)} &\ge& 1
\end{eqnarray}
as $N \rightarrow \infty$.
\end{lem}

\begin{proof}
Let's assume that $N$ is even. Suppose that we fix the first cluster at 0 and the $N$th cluster at $(N-1)d$. We define a vector interaction field between any two clusters as $F_{ij}=\alpha/{d_{ij}^{\eta+1}}$ with $\alpha = P_0/\eta$, and suppose that each cluster is denoted by a point on the interval $[0,(N-1)d]$ interacting with the others according to our vector field. Therefore, if we let the points be initially distributed on the line with distances $\{d_{ij}\}$, the final equilibrium configuration will be when $I_w(N,\{d_{ij}\})$ is minimized. This is because $I_w(\cdot,\cdot)$ acts as a potential function for the system due to the definition of the interaction vector field. Let $\{d_{ij}(f)\}$ denote the final distances of the clusters following equilibrium. If we consider $I_w(\cdot,\cdot)$ as a function of adjacent distances $\Delta_i \triangleq d_{i,{i+1}}$ for $i=1,\cdots,N-1$, it will have the following form:
\begin{equation}
I_w(N,\{\Delta_i\})=2 \sum_{i,j : j>i} \frac{P_0}{({\sum_i^j \Delta_i})^{\eta}}
\end{equation}
which is clearly a convex function of $\Delta_i$, $i=1,2,\cdots,N-1$. To find $\Delta_i$ for $i=1,2,\cdots,N-1$ in the equilibrium configuration, we need to minimize $I_w(N,\{\Delta_i\})$ subject to the constraint $\sum_{i=1}^{N-1} \Delta_i = Nd$. Since, the constraint is linear in $\Delta_i$ for $i=1,2,\cdots,N-1$ and the second derivative of $I_w(N,\cdot)$ with respect to any $\Delta_i$, $i=1,2,\cdots,N-1$ is a sum of some positive terms, there is no $i$, for which the second derivative of $I_w(\cdot,\cdot)$, the objective function, is zero. Therefore, the minimization has only one unique solution which is the equilibrium state (This is intuitively clear from the physical model of a number of points interacting on a one dimensional lattice). Since the system is in equilibrium under the distribution given by $\{d_{ij}(f)\}$, we have
\begin{equation}
D \Big( \frac{1}{N} I(N,\{\Delta_k\}) \Big)\Big|_{\Delta_{k}=\Delta_{k}(f)} = 0
\end{equation}
where $D(\cdot)$ denotes the exact differential and $\Delta_{k}(f) \triangleq d_{k,k+1}(f)$. We claim that if we choose $\Delta_k=d$ for all $k$, then for every $\epsilon$ small enough, we can find an $\tilde{N}$ so that for all $N > \tilde{N}$ we have:
\begin{equation}\label{limit}
\sum_i \frac{\partial \frac{1}{N} I(N, \{\Delta_{k}\})}{\partial \Delta_{i}}\Big |_{\Delta_{k}=\Delta_{k}(f)} < \epsilon \frac{\alpha}{d^{\eta+1}}
\end{equation}
This means that we can get arbitrarily close to the equilibrium point of the system. Since, the objective function is continuous and differentiable with respect to $\Delta_i$, for $i=1,2,\cdots,N-1$, we have chosen to approach the equilibrium point uniformly with respect to all $\Delta_i$, for $i=1,2,\cdots,N-1$, $\emph{i.e.}$, choosing the same amount of variation for all the coordinates. From the properties of a potential function for a vector field, we know that
\begin{equation}\label{f_i_1}
\sum_i \frac{\partial \frac{1}{N} I(N, \{\Delta_{k}\})}{\partial \Delta_{i}}\Big |_{\Delta_{k}=\Delta_{k}(f)} = \frac{1}{N} \sum_i F_i \le \frac{1}{N} \sum_i |F_i|
\end{equation}
where $F_i$ is the aggregate interaction field of all the other particles on the $i$th particle. If we set $\Delta_i=d$ for all the users, then
\begin{equation}\label{f_i_2}
|F_i| = \alpha \sum_{j=i+1}^{N-i+1} \frac{1}{(jd)^{\eta+1}} < \frac{\alpha}{d^{\eta+1}} \Big(\zeta(\eta+1) - \sum_{j=1}^i \frac{1}{j^{\eta+1}}\Big)
\end{equation}
for all $i \le N/2$, and $|F_i| = |F_{N-i+1}|$ for $i > N/2$. Therefore, we can bound the expression in the righthand side of Eq. (\ref{f_i_1}) as
\begin{equation}\label{f_i_3}
\frac{1}{N} \sum_i |F_i| < \frac{2}{N} \frac{\alpha}{d^{\eta+1}} \sum_{i=1}^{N/2}\Big(\zeta(\eta+1) - \sum_{j=1}^i \frac{1}{j^{\eta+1}}\Big) = \frac{\alpha}{d^{\eta+1}} \Big( \frac{2}{N} \sum_{i=1}^{N/2} a_i \Big)
\end{equation}
where $a_i \triangleq \zeta(\eta+1) - \sum_{j=1}^i \frac{1}{j^{\eta+1}}$. Clearly, $\lim_{i \rightarrow \infty} a_i = 0$. Therefore, if we choose $\tilde{N}$ so that $\sum_{i=\tilde{N}+1}^\infty \frac{1}{i^{\eta+1}} < \epsilon/2$, similar to the proof of Lemma \ref{zeta}, we can choose $N$ large enough to make sure that Eq. (\ref{limit}) holds. This means that the uniform spatial configuration asymptotically coincides with the global minimum of the system, which is the optimal spatial distribution of clusters with least aggregate interference. Since the result holds for large $N$, fixing the first and last cluster at $0$ and $(N-1)d$ does not affect the result. The case for odd $N$ can be treated similarly, since excluding one of the users does not change the asymptotic result.
\end{proof}
Lemma \ref{equilibrium} gives an important result: the worst-case aggregate interference of any spatial distribution of $N$ users in $[0,(N-1)d]$ is greater than or equal to that of a uniform linear array of $N$ clusters in $[0,(N-1)d]$, for sufficiently large $N$.

\begin{lem}\label{contrad}
Let $C_k(N,\{d_{ij}\})$ denote the set of clusters in frequency band $b_k$ corresponding to the optimal assignment strategy for a given spatial configuration of $N$ clusters in $[0,(N-1)d]$. Let $N_k(N,\{d_{ij}\})$ denote $|C_k(N,\{d_{ij}\})|$. Then, $\big\{N_k(N,\{d_{ij}\})\big\}_{N=1,2,\cdots}$ is an unbounded sequence, for all $k=0,\cdots,r-1$ and any spatial configuration of $N$ clusters in $[0,(N-1)d]$.
\end{lem}

\begin{proof}
Let $\mathcal{S}(N,\{d_{ij}\})=\{ N_k(N,\{d_{ij}\}) | k=0,\cdots,r-1\}$ and $N^*(N,\{d_{ij}\}) \triangleq \inf \mathcal{S}(N,\{d_{ij}\})$. Let's assume that $\{N^*(N,\{d_{ij}\})\}_{N=1,2,\cdots}$ is a bounded sequence. That is, there exist numbers $\tilde{N}$ and $M$ and a specific spatial configuration of the clusters for any $N > \tilde{N}$, such that $N^*(N,\{d_{ij}\}) < M$. Let $N^*(N,\{d_{ij}\})$ correspond to frequency band $b_{k^*}$ (whereas dependence on $N$ is implicit). We have
\begin{eqnarray}\label{r}
\frac{1}{N} I_o(r,N,\{d_{ij}\}) &=& \frac{1}{N} \sum_{k \neq k^*} \sum_{c_i,c_j \in C_k(N,\{d_{ij}\})} \frac{P_0}{d_{ij}^{\eta}} + \frac{1}{N} \sum_{c_i,c_j \in C_{k^*}(N,\{d_{ij}\})} \frac{P_0}{d_{ij}^{\eta}}
\end{eqnarray}
where we use the additional argument of $r$ to show the implicit dependence of the aggregate interference on the number of frequency bands. We take the same ensemble $c_i$ for $i=1,\cdots,N$ and assign any configuration of frequency bands $b_k$ for $k \neq k^*$ to all $c_i \in C_{k^*}(N,\{d_{ij}\})$ and keep the frequency assignment of the rest of the clusters as in the optimal strategy. Let $s_i$ denote the index corresponding to the new frequency band assigned to $c_i \in C_{k^*}(N,\{d_{ij}\})$. The normalized aggregate interference of this new ensemble will be
\begin{eqnarray}\label{r-1}
\frac{1}{N} I(r-1,N,\{d_{ij}\}) &=& \frac{1}{N} \sum_{k \neq k^*} \sum_{c_i,c_j \in C_k(N,\{d_{ij}\})} \frac{P_0}{d_{ij}^{\eta}}\\
\nonumber &+& \frac{1}{N} \sum_{k \neq k^*} \sum_{c_i \in C_{k^*}(N,\{d_{ij}\})} \sum_{c_j \in C_k(N,\{d_{ij}\})} \frac{P_0}{d_{ij}^{\eta}} \delta(s_i,k)
\end{eqnarray}

We also know that $I(r-1,N,\{d_{ij}\}) \ge I_o(r-1,N,\{d_{ij}\})$, where $I_o(r-1,N,\{d_{ij}\})$ is the aggregate interference of the optimal frequency band assignment to the same ensemble, using $r-1$ frequency bands. We define
\begin{equation}
g(N,\{d_{ij}\}) \triangleq \frac{1}{N^*(N,\{d_{ij}\})} \sum_{c_i,c_j \in C_{k^*}(N,\{d_{ij}\})} \frac{P_0}{d_{ij}^{\eta}}
\end{equation}
\begin{equation}
h(N,\{d_{ij}\}) \triangleq \frac{1}{N^*(N,\{d_{ij}\})} \sum_{k \neq k^*} \sum_{c_i \in C_{k^*}(N,\{d_{ij}\})} \sum_{c_j \in C_k(N,\{d_{ij}\})} \frac{P_0}{d_{ij}^{\eta}} \delta(s_i,k)
\end{equation}

Then, we have
\begin{eqnarray}\label{cont}
\frac{1}{N} I_o(r,N,\{d_{ij}\}) &=& \frac{1}{N} I(r-1,N,\{d_{ij}\}) + \frac{N^*(N,\{d_{ij}\})}{N} (g(N,\{d_{ij}\}) - h(N,\{d_{ij}\}))\\
\nonumber &\ge& \frac{1}{N} I_o(r-1,N,\{d_{ij}\}) + \frac{N^*(N,\{d_{ij}\})}{N} (g(N,\{d_{ij}\}) - h(N,\{d_{ij}\}))
\end{eqnarray}

Both $g(N,\{d_{ij}\})$ and $h(N,\{d_{ij}\})$ are upper bounded by $2 \zeta(\eta) \frac{P_0}{\delta^{\eta}}$, where $\delta$ is the minimum distance between two clusters, based on our assumptions. We have $\frac{N^*(N,\{d_{ij}\})}{N} \big| (g(N,\{d_{ij}\}) - h(N,\{d_{ij}\})) \big| < 4 \zeta(\eta) \frac{P_0}{\delta^{\eta}} \big(\frac{M}{N}\big)$. Therefore, we can choose $N$ sufficiently large with a specific corresponding ensemble, in such a way that $\frac{N^*(N,\{d_{ij}\})}{N} |g(N,\{d_{ij}\}) - h(N,\{d_{ij}\})|$ becomes arbitrarily small. Equation (\ref{cont}) states that the normalized aggregate interferences using $r$ frequency bands and $r-1$ frequency bands for a specific configuration may become arbitrarily close.

This is clearly not possible. Suppose that we have $r-1$ frequency bands $b_0,b_1,\cdots,b_{r-2}$. Consider the optimal frequency band assignment to the clusters $c_i$ for $i=1,2,\cdots,N$ in $[0,(N-1)d]$. Since $\sum_k N_k(N,\{d_{ij}\}) = N$, there exists a $0 \le  k_0 \le r-2$ such that $N_{k_0}(N,\{d_{ij}\}) \ge N/(r-1)$. If we consider the clusters in $C_{k_0}(N,\{d_{ij}\})$ and allow the assignment of an additional frequency band $b_{r-1}$ to them, then according to Theorem \ref{upper}, there exists a frequency band assignment using the two bands $b_{k_0}$ and $b_r$, to the clusters in $C_{k_0}(N,\{d_{ij}\})$, for which the normalized aggregate interference of all clusters in $C_{k_0}(N,\{d_{ij}\})$ is at least half of that in the worst case scenario. Therefore, adding a frequency band $b_{r-1}$, decreases the normalized aggregate interference by at least
\begin{equation}
\frac{1}{2 N} \sum_{c_i,c_j \in C_{k_0}(N,\{d_{ij}\})} \frac{P_0}{d_{ij}^{\eta}} \ge \frac{1}{(r-1)^{\eta+1}} \zeta(\eta) \frac{P_0}{d^{\eta}}
\end{equation}
for sufficiently large $N$, according to Lemma \ref{equilibrium}. Therefore, $\frac{1}{N} I_o(r-1,N,\{d_{ij}\})$ can not be arbitrarily close to $\frac{1}{N} I_o(r,N,\{d_{ij}\})$ for any spatial configuration of clusters, as $N \rightarrow \infty$. Thus, $\{N^*(N,\{d_{ij}\})\}_{N=1,2,\cdots}$ is an unbounded sequence. Since $N^*(N,\{d_{ij}\}) = \inf \mathcal{S}(N,\{d_{ij}\})$ and $\mathcal{S}(N,\{d_{ij}\})$ is a finite set, we conclude that $\{N_k(N,\{d_{ij}\})\}_{N=1,2,\cdots}$ is an unbounded sequence for all $k=0,1,\cdots,r-1$.
\end{proof}

\begin{lem}\label{lower_partial}
Let $N_k$ denote $|C_k|$ corresponding to the optimal strategy for a given spatial configuration of clusters in $[0,(N-1)d]$. Then, we have $\frac{I_w(N_k,\{d_{ij}\})|_{c_i,c_j \in C_k}}{I_w(N_k, \frac{N}{N_k}d)} \ge 1$ for all $k$, for sufficiently large $N$.
\end{lem}

\begin{proof}
Using the result of Lemma \ref{equilibrium}, we can write
\begin{equation}
\frac{1}{N_k} I_w(N_k,\{d_{ij}\})|_{c_i,c_j \in C_k} \ge \frac{1}{N_k} I_w(N_k, \frac{N}{N_k}d)
\end{equation}
since the clusters are located in $[0,(N-1)d]$ and $N_k$ is unbounded for all $k$ (Lemma \ref{contrad}) as $N \rightarrow \infty$.\\
\end{proof}

\emph{Proof of Theorem \ref{optimal_strategy}}:
We have
\begin{eqnarray}
\lim_{N \rightarrow \infty} \frac{1}{N} I_o(N,\{d_{ij}\}) &=& \lim_{N \rightarrow \infty} \frac{1}{N} \sum_{k=0}^{r-1} \sum_{c_l, c_m \in C_k} I_w(N_k,\{d_{lm}\})\\
\nonumber &\ge& \lim_{N \rightarrow \infty} \frac{1}{N} \sum_{k=0}^{r-1} I_w(N_k, \textstyle \frac{N}{N_k}d) \displaystyle \ge \lim_{N \rightarrow \infty} {\frac{\sum_{k=0}^{r-1} N_k^{\eta+1}}{N^{\eta+2}}} I_w(N,d)
\end{eqnarray}
where we have used Lemmas \ref{zeta}, \ref{contrad} and \ref{lower_partial}. The expression $\sum_{k=0}^{r-1} N_k^{\eta+1}$ achieves its minimum when $N_k = N/r$, for all $k=0,\cdots,r-1$. Therefore,
\begin{eqnarray}
\lim_{N \rightarrow \infty} \frac{1}{N} I_o(N,\{d_{ij}\}) \ge \lim_{N \rightarrow \infty} {\frac{\sum_{k=0}^{r-1} N_k^{\eta+1}}{N^{\eta+2}}} I_w(N,d) \ge \frac{1}{r^{\eta}} \lim_{N \rightarrow \infty} \frac{1}{N} I_w(N,d) = \frac{1}{r^{\eta}} 2 \zeta(\eta) \frac{P_0}{d^{\eta}}
\end{eqnarray}
which gives the statement of Theorem \ref{optimal_strategy}.

\subsection{Proof of Theorem \ref{regular}}\label{regular_a}

\begin{lem}\label{sandwich}
Let $I_o(N,d) \triangleq I_o(N,\{d_{ij}\})$ and $I_w(N,d) \triangleq I_w(N,\{d_{ij}\})$ denote the aggregate interference for the optimal alternating strategy and worst-case scenario, for a uniform linear array in $[0,(N-1)d]$. Then,
\begin{center}
$\textstyle r I_w(\lfloor N/r \rfloor,rd) \le I_o(N,d) \le r I_w(\lceil N/r \rceil,rd)$
\end{center}
\end{lem}

\begin{proof}
We have a uniform linear array of $N$ clusters in $[0,(N-1)d]$. In case of the alternating frequency assignment, the clusters corresponding to the $b_k$ band make a uniform linear array of $\lfloor N/r \rfloor \le N_k \le \lceil N/r \rceil$ clusters each distance $rd$ apart from its neighbor(s), for all $k=0,\cdots,r-1$. Therefore, the aggregate interference is equivalent to the sum of aggregate interferences of the $r$ disjoint sets of users corresponding to the $r$ frequency bands. Clearly, all the clusters in each subset have the same frequency band and therefore, the aggregate interference for each subset will be $I_w(N_k,rd)$. We can write $I_o(N,d)$ as $I_o(N,d)=\sum_{k=1}^{r-1} I_w(N_k,rd)$. Since $I_w(N,\cdot)$ is an increasing function of $N$, we can sandwich $I_w(N_k,rd)$ as
\begin{equation}
I_w(\lfloor N/r \rfloor,rd) \le I_w(N_k,rd) \le I_w(\lceil N/r \rceil,rd)
\end{equation}

Thus,
\begin{equation}
r I_w(\lfloor N/r \rfloor,rd) \le \sum_{k=0}^{r-1} I_w(N_k,rd) \le r I_w(\lceil N/r \rceil,rd)
\end{equation}
which gives the statement of the Lemma.
\end{proof}

%\begin{cor}
%In D dimensions, for a lattice of size $N^D$ and distance $d$, as $N \rightarrow \infty$
%
%\begin{center}
%$\frac{1}{N^D} I_w(N,d) \rightarrow \frac{A(\eta,D)}{d^{\eta}}$
%\end{center}
%where $A(\eta,D) = 2^D \sum_{i_1,i_2,\cdots,i_D=1}^\infty \frac{1}{(i_1^2 + i_2^2 + \cdots + i_D^2)^{\eta/2}}$ is independent of $N$.
%\end{cor}
%
%\begin{proof}
%The proof is a straightforward generalization of the one dimensional case.\\
%\end{proof}

\emph{Proof of Theorem \ref{regular}}:

Using Lemmas \ref{sandwich} and \ref{zeta}, and using the Sandwich Theorem in Calculus we can see that,
\begin{eqnarray}
\lim_{N \rightarrow \infty} \frac{1}{N} I_o(N,d) = \lim_{N \rightarrow \infty} \frac{1}{N} r I_w(\lceil N/r \rceil,rd) = \lim_{N \rightarrow \infty} \frac{\lceil N/r \rceil}{N/r} 2 \zeta(\eta) \frac{P_0}{(rd)^{\eta}} = \frac{1}{r^{\eta}} 2 \zeta(\eta) \frac{P_0}{d^{\eta}}
\end{eqnarray}
from which the statement of the Theorem follows.
\subsection{Proof of Theorem \ref{r=2}}\label{optimal_2}
\begin{lem}\label{three}
Let's consider a specific assignment of the frequency bands to $N$ clusters in a Uniform Linear Array, such that there are 3 successive co-band clusters. Switching the frequency band of the middle cluster decreases the aggregate interference, regardless of the frequency bands of the rest of the clusters, if $\zeta(\eta) < 2$, where $\zeta(\cdot)$ is the Riemann zeta function.
\end{lem}

\begin{proof}
Without loss of generality, assume that $c_{i-1}$, $c_i$ and $c_{i+1}$ are in frequency band $b_1$. The interference of all the other clusters on $c_i$ is clearly greater than or equal to $2P_0/d^{\eta}$. If it switches to band $b_0$, the interference of all the other clusters on $c_i$ will be less than or equal to $2(\zeta(\eta)-1)P_0/d^{\eta}$, for which the equality occurs when there are infinite number of users other than $c_{i-1}$ and $c_{i+1}$ in band $b_0$. Therefore, the aggregate interference will decrease after switching if $2(\zeta(\eta)-1) < 2$, which gives the statement of the Lemma.
\end{proof}

\begin{lem}\label{partition}
For any given frequency assignment, we can find a configuration which has smaller aggregate interference, and can be partitioned into subsets of clusters with the property that in each subset all the clusters have alternating frequency assignment.
\end{lem}

\begin{proof}
Using Lemma \ref{three}, we can break any 3 or higher successive co-band clusters into at most 2 successive co-band clusters, by repeatedly switching the frequency band of the middle cluster. We will then get a configuration which has smaller aggregate interference compared to the original one (according to Lemma \ref{three}). We can therefore partition this new configuration into subsets of the stated property, by just putting the boundaries midway between every two successive co-band clusters, \emph{i.e.}, if we have $k$ pairs of co-band clusters, we can partition the clusters into $k+1$ subsets $S_1,S_2,\cdots,S_{k+1}$, in such a way that all the clusters in $S_i$ have alternating frequency assignment, for all $i=1,2,\cdots,k+1$.
\end{proof}

\begin{lem}\label{reduction}
Let's consider the subsets $S_1,S_2,\cdots,S_{k+1}$ given by the partitioning procedure in Lemma \ref{partition}. Then, switching all the frequency bands in $S_1$, decreased the aggregate interference.
\end{lem}

\begin{proof}
Without loss of generality, let's assume that there are $N_{S_1}=2l$ clusters in $S_1$ and the first cluster is in band $b_0$. Since $N_{S_1}$ is even, the last cluster in $S_1$ is in band $b_1$. Also, the first cluster in $S_2 \bigcup S_3 \bigcup \cdots \bigcup S_{k+1}$ is in band $b_1$. Let $I_1(b)$ and $I_1(a)$ respectively denote the interference of all the clusters in $C \backslash S_1$ on the clusters in $S_1$ in band $b_1$, before and after switching. Therefore,
\begin{equation}
\Delta I_1 \triangleq I_1(b) - I_1(a) \ge \frac{P_0}{d^{\eta}} \textstyle (1 -\frac{1}{2^{\eta}} + \frac{1}{3^{\eta}} - \frac{1}{4^{\eta}} + \cdots + \frac{1}{(2l-1)^{\eta}} - \frac{1}{(2l)^{\eta}})
\end{equation}

Similarly, let $I_0(b)$ and $I_0(a)$ respectively denote the interference of all the clusters in
$C \backslash S_1$ on the clusters in $S_1$ in band $b_0$, before and after switching. Therefore,
\begin{eqnarray}
|\Delta I_0| \triangleq |I_0(b) - I_1(a)| &\le& \frac{P_0}{d^{\eta}} \textstyle \big \{(\zeta(\eta) - 1) - (\zeta(\eta) - 1 - \frac{1}{2^{\eta}}) + \textstyle (\zeta(\eta) - 1 - \frac{1}{2^{\eta}} - \frac{1}{3^{\eta}})\\
\nonumber &-& \textstyle (\zeta(\eta) - 1 - \frac{1}{2^{\eta}} - \frac{1}{3^{\eta}} - \frac{1}{4^{\eta}}) + \cdots - \textstyle (\zeta(\eta) - 1 - \frac{1}{2^{\eta}} - \cdots - \frac{1}{(2l)^{\eta}}) \big \}\\
\nonumber &=& \frac{P_0}{d^{\eta}} \Big(\frac{1}{2^{\eta}} + \frac{1}{4^{\eta}} + \cdots + \frac{1}{(2l)^{\eta}}\Big)
\end{eqnarray}

Therefore, the difference in the aggregate interference of the system before and after switching will be
\begin{eqnarray}
\Delta I \triangleq 2(\Delta I_1 + \Delta I_0) &\ge& 2\frac{P_0}{d^{\eta}} \textstyle \big \{ (1+\frac{1}{3^{\eta}}+\cdots+\frac{1}{(2l-1)^{\eta}}) - \frac{2}{2^{\eta}} (1 + \frac{1}{2^{\eta}} + \cdots + \frac{1}{l^{\eta}}) \big \}\\
\nonumber &>& 2\frac{P_0}{d^{\eta}} \big \{ 1 - \frac{2}{2^{\eta}} \zeta(\eta) \big \}
\end{eqnarray}

$1 - \frac{2}{2^{\eta}} \zeta(\eta)$ is clearly positive for all $\eta \ge 2$. This means that switching all the frequency bands in $S_1$ decreases the aggregate interference. $\eta \ge 2$ also satisfies the bound in Lemma \ref{three}. Thus, the optimal solution is the alternating solution for $\eta \ge 2$. The case for odd $N$ can be treated in a similar manner.
\end{proof}

\emph{Proof of Theorem \ref{r=2}}:

Suppose that we are given a specific frequency band assignment for $N$ clusters in a Uniform Linear Array. Using Lemma \ref{partition}, we can partition the clusters into subsets $S_1,S_2,\cdots,S_{k+1}$. Then, using Lemma \ref{reduction}, we switch all the frequency bands in $S_1$. After switching, the last cluster in $S_1$ will not be co-band with the first cluster in $S_2$. Therefore, we can combine $S_1$ and $S_2$ to get $S'_2=S_1 \bigcup S_2$, reducing the number of partitions to $k$. We can now switch all the frequency bands in $S'_2$ and combine it with $S_3$ to get $S'_3=S'_1 \bigcup S_2 = S_1 \bigcup S_2 \bigcup S_3$. If we repeat this procedure a total of $k$ times, the resulting configuration is clearly an alternating assignment. Since the aggregate interference decreases in each of the $k$ steps, the final alternating configuration has smaller interference than the initial configuration. Noting that the initial configuration was arbitrary, we conclude that the alternating frequency assignment strategy is the optimal strategy of frequency assignment for a finite Uniform Linear Array with $r=2$ and $\eta \ge 2$.

\subsection{Proof of Corollary \ref{c1}}\label{bound_1}
Clearly, we have
$I_w(N,d_{max}) \le I_w(N,\{d_{ij}\}) \le I_w(N,d_{min})$, where $d_{min}$ and $d_{min}$ are the minimum and maximum spacing between the clusters, respectively. The statement of Theorem \ref{optimal_strategy}, can be written as
\begin{equation}\label{assymp_best_2}
\frac{I_o(N,\{d_{ij}\})}{I_w(N,d_{min})} \ge \frac{1}{r^{\eta}} \Big(\frac{d_{min}}{d}\Big)^{\eta}
\end{equation}
as $N \rightarrow \infty$. We also have $I_o(N,\{d_{ij}\}) \ge I_o(N,d_{max})$, where $I_o(N,d_{max})$ is the aggregate interference of the optimal frequency band allocation for a Uniform Linear Array in $[0,(N-1)d_{max}]$. Using Theorem \ref{regular}, we have
\begin{equation}\label{assymp_best_4}
\lim_{N \rightarrow \infty} \frac{I_o(N,d_{max})}{I_w(N,d_{min})} = \frac{1}{r^{\eta}}
\Big(\frac{d_{min}}{d_{max}}\Big)^{\eta}
\end{equation}

Combining Eqs. (\ref{assymp_best_2}) and (\ref{assymp_best_4}) we get
\begin{equation}\label{lower_opt}
\frac{I_o(N,\{d_{ij}\})}{I_w(N,d_{min})} \ge \frac{1}{r^{\eta}} \Big(\frac{d_{min}}{\min\{d_{max},d\}}\Big)^{\eta}
\end{equation}

Using Theorem \ref{upper}, a bound of
$I_a(N,\{d_{ij}\}) \le \frac{1}{r} I_w(N,\{d_{ij}\}) \le \frac{1}{r}
I_w(N,d_{min})$ is obvious, where $I_a(N,\{d_{ij}\})$ is the aggregate interference corresponding to the output of the Main Algorithm. Combining this bound with Eq. (\ref{lower_opt}), we get
\begin{equation}
\frac{I_a(N,\{d_{ij}\})}{I_o(N,\{d_{ij}\})} \le \frac{r^{\eta-1}}{\big(\frac{d_{min}}{\min\{d_{max},d\}}\big)^{\eta}}
\end{equation}
as $N \rightarrow \infty$. The latter gives the statement of the Corollary.

\section{Dynamics Of The Interference}

\subsection{No Time Variation In The Activity of Clusters}\label{dynamic1}

We consider the case of two accessible frequency bands, $b_0$ and $b_1$, for mathematical convenience. Generalization to more than two frequency bands is straightforward. We associate $\epsilon_i=-1$ and $\epsilon_i=1$ to clusters in band $b_0$ and $b_1$, respectively.

If $c_i$ is active at time $t$, it will experience an interference

\begin{equation}
I_{c_i}(t) = \frac{1}{2}\bigg(I_i + \epsilon_i(t) \sum_{j \neq i} \frac{\epsilon_j(t)}{d_{ij}^\eta} \bigg)
\end{equation}
where $I_i$ is the worst case interference experienced by cluster $c_i$.

We define the band $b_j$ to be \emph{appropriate} for cluster $c_i$, if $c_i$ is assigned in band $b_j$ in the target configuration of the algorithm. If $c_i$ is not assigned in its appropriate band at time $t$, it will increase the aggregate interference by

\begin{equation}
2 \Delta I_{c_i}(t) = 2 \Big|  \sum_{j \neq i} \frac{\epsilon_j(t)}{d_{ij}^\eta} \Big|
\end{equation}

If a total of $M(t)$ users $c_i$, $i \in \{ i_1,\cdots,i_{M(t)} \}$ are not in the \emph{appropriate} frequency bands, the aggregate interference will be
\begin{equation}
I(t)=I_a + \sum_{k=1}^{M(t)} 2 \Big|\sum_{j \neq i_k} \frac{\epsilon_{j}(t)}{d_{i_k, j}^\eta}\Big|
\end{equation}
where $I_a$ is the target performance of the algorithm.

Assuming spatial ergodicity, we can average over an ensemble of different update patterns. Thus,

\begin{equation}\label{diff}
\mathcal{E} [I(t)]- \mathcal{E} [I_a] = 2M(t) \mathcal{E}\Big[\Big|\sum_{j \neq i} \frac{\epsilon_{j}(t)}{d_{i j}^\eta}\Big|\Big]
\end{equation}
where $\mathcal{E}$ denotes the ensemble average.

After any update, the average change in $\mathcal{E}[I(t)]$, will be
\begin{equation}\label{change}
\Delta \mathcal{E} [I(t)] = \frac{\rho M(t)}{N} 2\mathcal{E}\Big[\big|\sum_{j \neq i} \frac{\epsilon_{j}(t)}{d_{i j}^\eta}\big|\Big]
\end{equation}
where $\rho$ is a geometrical constant showing the effective number of interacting neighbors to a cluster including itself. We are assuming that every cluster which is not in the appropriate band, increases the interference on an effective number of its neighbors (this is a linearization near the equilibrium point).

Combining Eqs. (\ref{diff}) and (\ref{change}), we get
\begin{equation}
\frac{\Delta \mathcal{E} [I(t)]}{\Delta T} = -\frac{\rho}{N \Delta T} (\mathcal{E}[I(t)] - \mathcal{E}[I_a])
\end{equation}
where $\Delta T$ is the inverse update rate. On the time scale of the updates, using the ansatz $\tau \triangleq N \Delta T$ one can write
\begin{equation}
\frac{d \mathcal{E}[{I}(t)]}{dt} = -\frac{\rho}{\tau} (\mathcal{E} [I(t)] - \mathcal{E}[I_a])
\end{equation}

\subsection{The Variance Equation Under Time-varying Activity of Clusters}\label{dynamic2}

Using the It\^{o} rule, the variance equation associated with Eq. (\ref{ito}) is
\begin{eqnarray}
d \mathcal{I}^2(t) &=& -\frac{2 \rho}{\tau} \mathcal{I}(t) (\mathcal{I}(t) - \mathcal{I}_a(t)) dt\\
\nonumber &+& \bigg( \Big(\mathcal{I}(t) + \frac{4}{N}\mathcal{I}(t)\Big)^2 - \mathcal{I}^2(t)\bigg) dN_+ + \bigg( \Big(\mathcal{I}(t) - \frac{4}{N}\mathcal{I}(t)\Big)^2 - \mathcal{I}^2(t)\bigg) dN_-
\end{eqnarray}

By simplifying the expression and taking expectations we get
\begin{equation}
\frac{d E(\mathcal{I}^2(t))}{d t} = -\Big(\frac{2\rho}{\tau} - \frac{32 \lambda}{N^2}\Big) E({\mathcal{I}^2(t)}) + \frac{2 \rho}{\tau} \mathcal{I}^2_a
\end{equation}

Therefore, in the steady state the variance settles down to
\begin{eqnarray}
\sigma_{ss}^2 &\triangleq& E\Big[\big(\mathcal{I}(t) - E[\mathcal{I}(t)]\big)^2\Big] = \mathcal{I}^2_a \frac{2\rho / \tau}{2\rho/\tau - 32 \lambda/N^2} - \mathcal{I}^2_a = \mathcal{I}^2_a \frac{16 \lambda \tau}{N^2 \rho - 16 \lambda \tau}
\end{eqnarray}
given
\begin{equation}
\frac{16 \lambda \tau}{N^2 \rho} < 1
\end{equation}

\begin{figure} \label{model}
\begin{center}
    \includegraphics[width=3in]{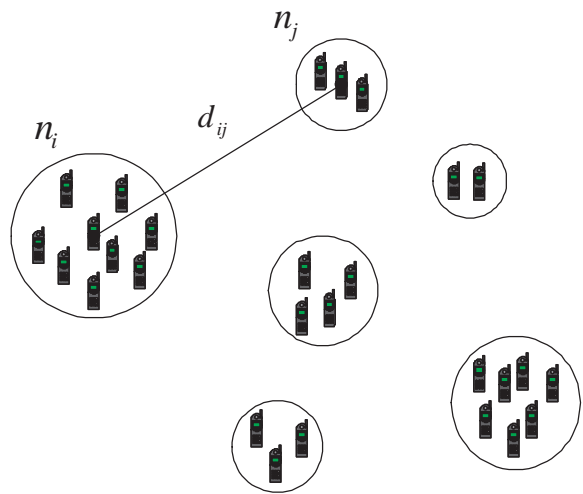}
\end{center}
\caption{System Model}
\end{figure}

\begin{figure}[h!]
\begin{center}
$\begin{array}{c@{\hspace{.01in}}c@{\hspace{.01in}}c}
\includegraphics[width=2.9in]{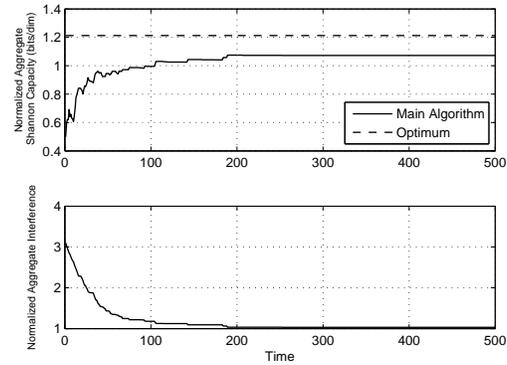}\\
\includegraphics[width=2.9in]{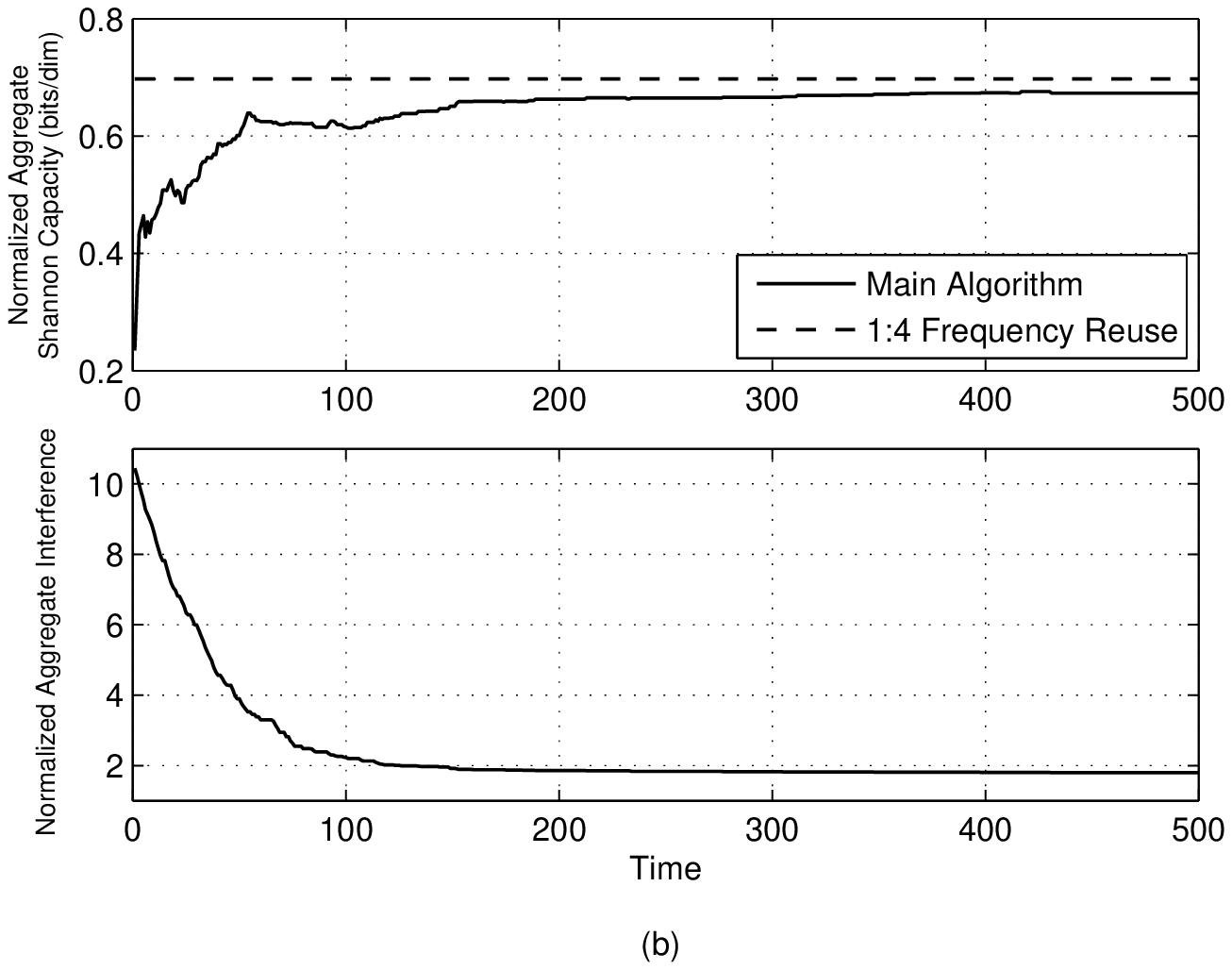}  \\
\includegraphics[width=2.9in]{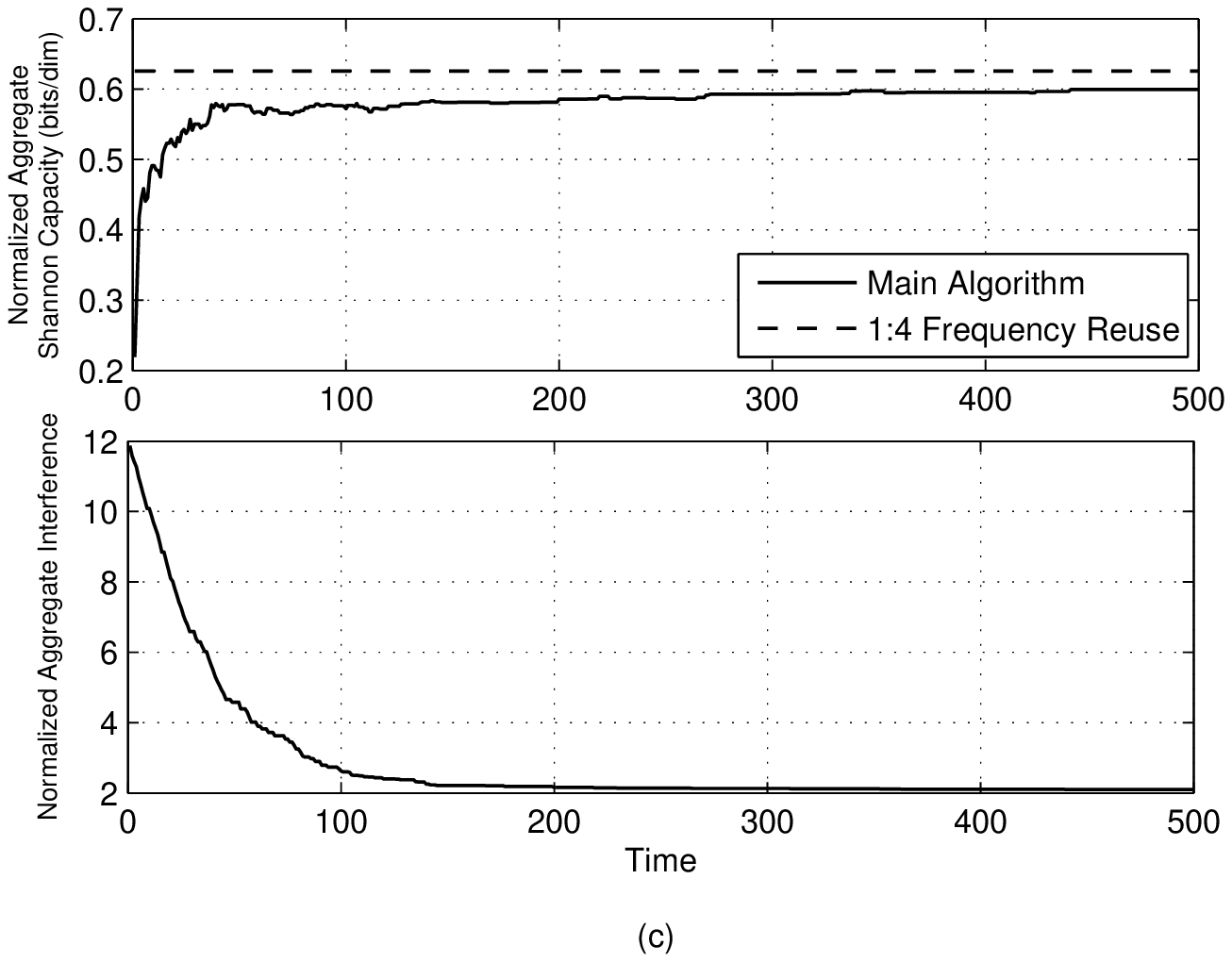}  \\
\end{array}$
\end{center}
\caption{\scriptsize Normalized aggregate Shannon capacity and normalized aggregate interference curves for arrays of 100 clusters vs. time, (a) uniform linear array, (b) rectangular lattice, and (c) hexagonal lattice.}
\label{fig:contoursStat}
\end{figure}

\begin{figure}
\begin{center}
\includegraphics[width=3.2in]{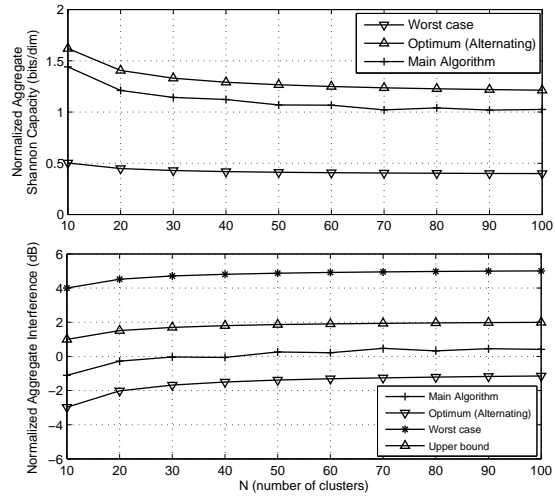}
\end{center}
\caption{\scriptsize Normalized aggregate Shannon capacity and normalized aggregate interference curves for uniform linear arrays vs. N}
\end{figure}

\begin{figure}[h!]
\begin{center}
$\begin{array}{c@{\hspace{.01in}}c@{\hspace{.01in}}c}
\includegraphics[width=3.2in]{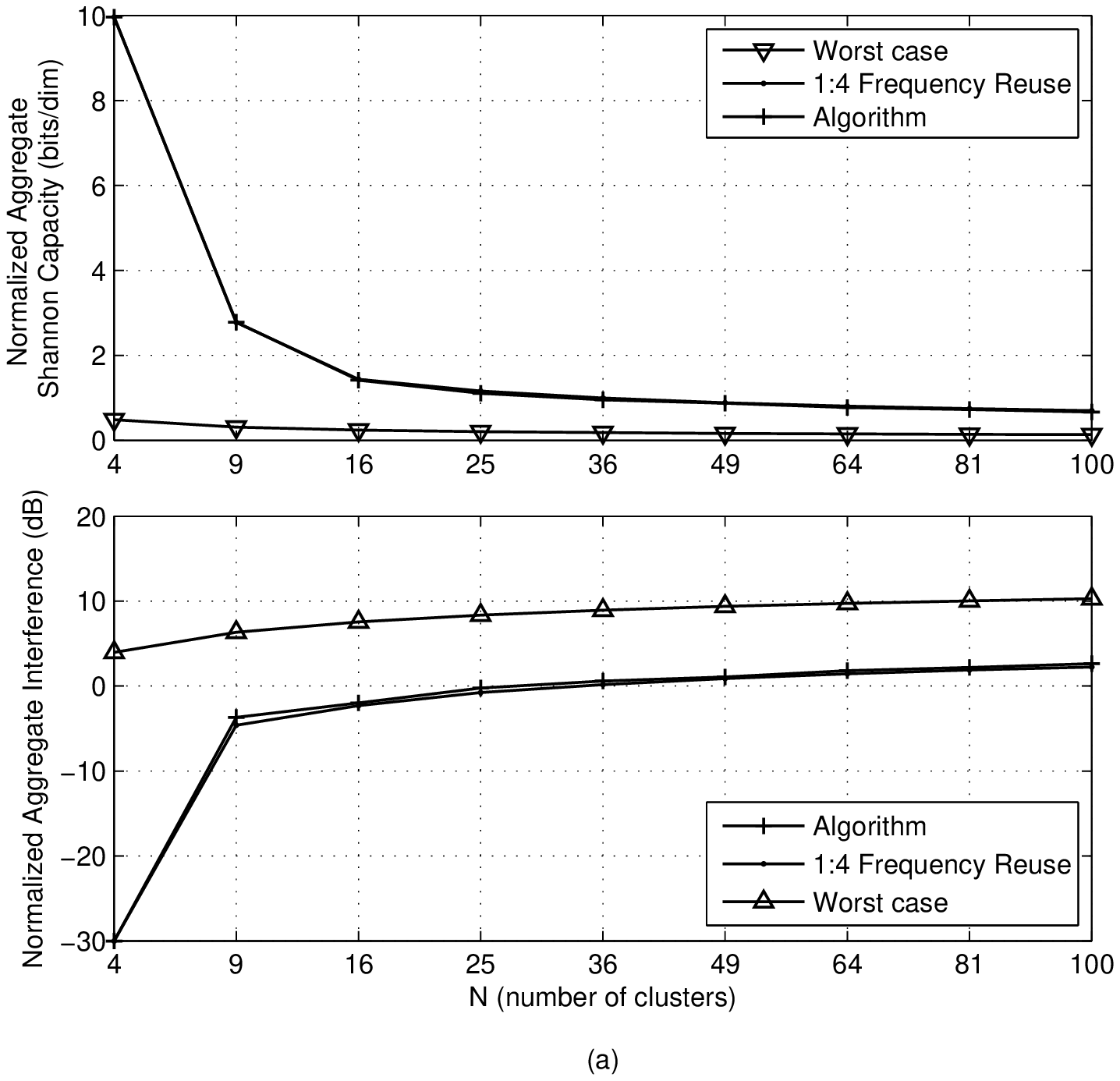} &
\includegraphics[width=3.2in]{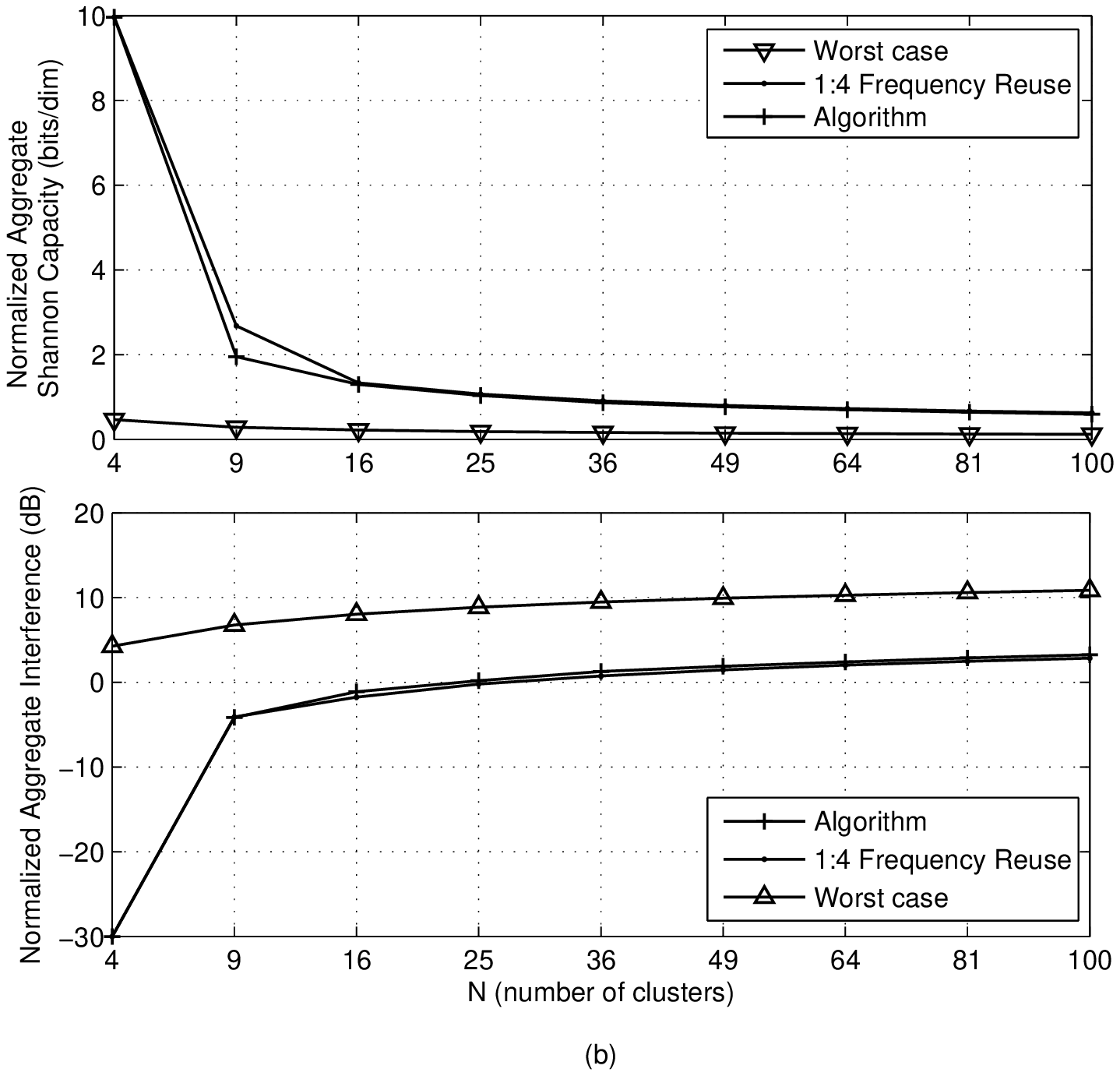}
\end{array}$
\end{center}
\caption{\scriptsize Normalized aggregate Shannon capacity and normalized aggregate interference performance curves for 2 dimensional arrays vs. N. (with $r=4$) (a) rectangular lattice, and  (b) hexagonal lattice}
\label{fig:contoursStat}
\end{figure}

\begin{figure}
\begin{center}
\includegraphics[width=3.6in]{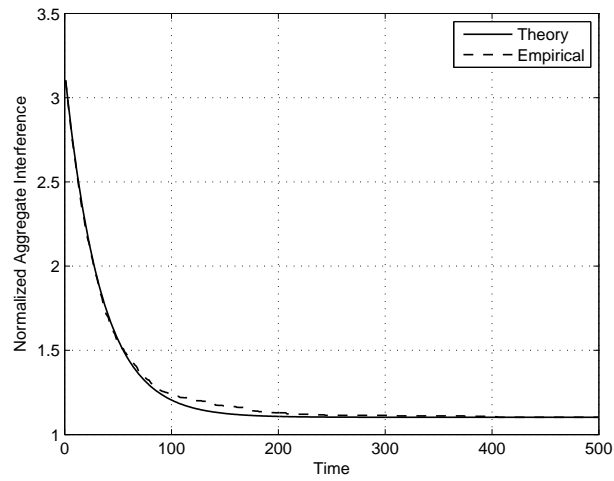}
\end{center}
\caption{Normalized aggregate interference vs. time for a uniform linear array of 100 clusters with $\alpha=1$.}
\end{figure}

\begin{figure}
\begin{center}
\includegraphics[width=3.6in]{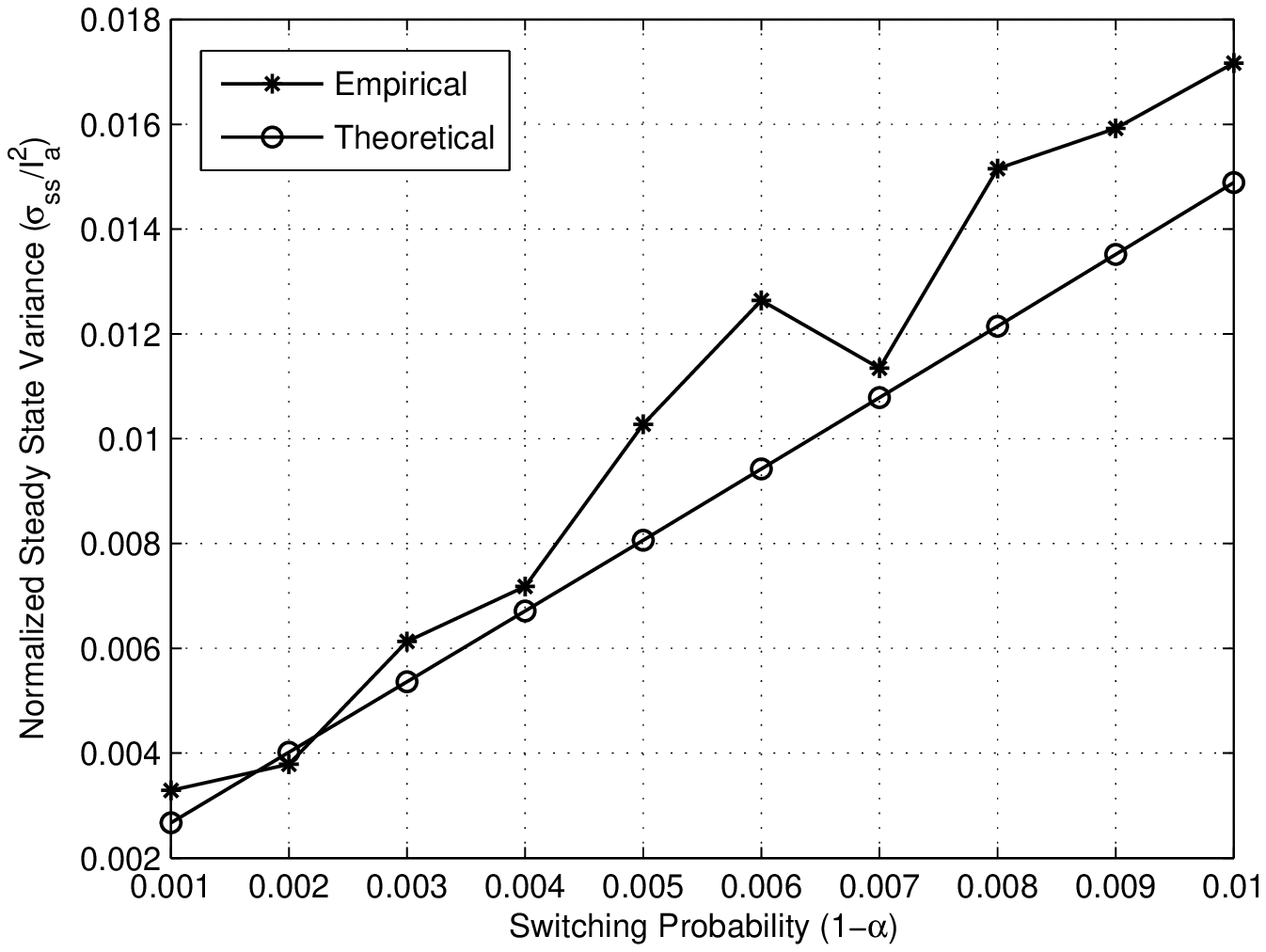}
\end{center}
\caption{Normalized steady state variance vs. switching rate for a uniform linear array of 100 clusters.}
\end{figure}


\begin{thebibliography}{99}

\bibitem{collision} B. Babadi, V. Tarokh, ``The Impact Of Scanning Time On Distributed Dynamic Frequency Allocation'', preprint, to be presented in Asilomar Conference on Signals, Systems and Computers, 2008.

\bibitem{cendrillon} R. Cendrillon, J. Huang, M. Chiang, and M. Moonen, ``Autonomous Spectrum Balancing for Digital Subscriber Lines'', IEEE Transactions on Signal Processing, Vol. 55, No. 8, August 2007.

\bibitem{sloane} J. H. Conway and N. J. A. Sloane, \emph{Sphere Packings, Lattices and Groups}, Springer-Verlag NY, First Edition, 1988.

\bibitem{etkin} R. Etkin, A. Parekh, and D. Tse, ``Spectrum Sharing for Unlicensed Bands'', IEEE Journal On Selected Areas In Communications, Vol. 25, No. 3, April 2007.

\bibitem{feller} W. Feller, \emph{Probability Theory}, Vol. I, John Wiley, New York, 1957 and Vol. II, 1966.

\bibitem{huang} J. Huang, R. A. Berry, and M. L. Honig, ``Distributed Interference Compensation for Wireless Networks'', IEEE Journal On Selected Areas In Communications, Vol. 24, No. 5, May 2006.

\bibitem{Ito} K. It\^{o}, ``Stochastic Integral'', Proc. Imperial Acad. Tokyo, Vol. 20, pp. 519-524, 1994.

\bibitem{macdonald} V. H. MacDonald, ``The Cellular Concept'', The Bell Systems Technical Journal, Vol. 58, No. 1, pp. 15-43, Januaray 1979.

\bibitem{peng} C. Peng, H. Zheng, B. Y. Zhao, ``Utilization and Fairness in Spectrum Assignment for Opportunistic Spectrum Access'', ACM Monet, May 2006.

\bibitem{ramanathan} S. Ramanathan, ``A Unified Framework and Algorithm for Channel Assignment in Wireless Networks'', Wireless Networks, Volume 5, Issue 2, March 1999.

\bibitem{steele} R. Steele, L. Hanzo, \emph{Mobile Radio Communications}, Wiley-IEEE Press, August 1999.

\bibitem{yu} W. Yu, G. Ginis, and J. M. Cioffi, ``Distributed Multiuser Power Control for Digital Subscriber Lines'', IEEE Journal On Selected Areas In Communications, Vol. 20, No. 5, June 2002.
\end{thebibliography}
\end{document}